# Attosecond Electron Pulse Trains and Quantum State Reconstruction in Ultrafast Transmission Electron Microscopy

Katharina E. Priebe[1], Christopher Rathje[1], Sergey V. Yalunin[1], Thorsten Hohage[2], Armin Feist[1], Sascha Schäfer[1], and Claus Ropers[1,3,*]

[1] 4th Physical Institute – Solids and Nanostructures, University of Göttingen, Germany

[2] Institut für Numerische und Angewandte Mathematik, University of Göttingen, Germany

[3] International Center for Advanced Studies of Energy Conversion (ICASEC), University of Göttingen, Germany

**Abstract**

*We introduce a framework for the preparation, coherent manipulation and characterization of free-electron quantum states, experimentally demonstrating attosecond pulse trains for electron microscopy. Specifically, we employ phase-locked single-color and two-color optical fields to coherently control the electron wave function along the beam direction. We establish a new variant of quantum state tomography –"SQUIRRELS" – to reconstruct the density matrices of free-electron ensembles and their attosecond temporal structure. The ability to tailor and quantitatively map electron quantum states will promote the nanoscale study of electron-matter entanglement and the development of new forms of ultrafast electron microscopy and spectroscopy down to the attosecond regime.*

Optical, electron and x-ray microscopy and spectroscopy reveal specimen properties via spatial and spectral signatures imprinted onto a beam of radiation or electrons. Leaving behind the traditional paradigm of idealized, simple probe beams, advanced optical techniques increasingly harness tailored probes, or even their quantum properties and probe-sample entanglement. The rise of structured illumination microscopy[1], pulse shaping[2], and multidimensional[3] and quantum-optical spectroscopy[4] exemplify this development. Similarly, electron microscopy explores the use of shaped electron beams exhibiting particular spatial symmetries[5] or angular momentum[6,7], and novel measurement schemes involving quantum aspects of electron probes have been proposed[8,9]. Ultrafast imaging and spectroscopy with electrons and x-rays are the basis for an ongoing revolution in the understanding of dynamical processes in matter on atomic scales[10–13]. The underlying technology heavily rests on laser science for the generation and characterization of ever-shorter femtosecond electron[10,14] and x-ray[15–17] probe pulses, with examples in optical pulse compression[18] and streaking spectroscopy[19–21]. The temporal structuring of electron probe beams is facilitated by time-dependent fields in the radio-frequency[22–24], terahertz[18,25] or optical domains. Promising a further leap in temporal resolution, recent findings suggest that ultrafast electron diffraction and microscopy with optically phase-controlled and sub-cycle, attosecond-structured wave functions may be feasible[8,26–30]. Specifically, light-field control may translate the temporal resolution of ultrafast transmission electron microscopy (UTEM)[31,32] and electron diffraction (UED)[10,33], currently at about 200 fs[34] and 20 fs[14,23], respectively, to the range of attoseconds[26,27,35]. However, such future technologies call for means to both prepare and fully analyze the corresponding quantum states of free electrons.

*Email: claus.ropers@uni-goettingen.de

Here, we demonstrate the coherent control and attosecond density modulation of free-electron quantum states using multiple phase-locked optical interactions. Moreover, we introduce quantum state tomography for free electrons, providing crucial elements for ultrafast free-electron quantum optics. In the first set of experiments, (sketched in Fig. 1**a**), two laser beams at frequencies $\omega$ and $2\omega$ are focused onto a single-crystalline graphite flake that is transparent for 120-keV electrons. A pulsed electron beam, generated by an ultrafast field-emission cathode[34], traverses the dual-color optical near-field, and its kinetic energy spectrum is subsequently recorded. The relative phase between the two laser pulses is precisely controlled by a pair of dispersive wedges. Single-color excitation (upper two panels in Fig. 1**c**) induces spectra with symmetric sideband peaks separated by the respective photon energy, as previously reported in the context of photon-induced near-field electron microscopy (PINEM)[36,29,30,26,37] and free-electron Rabi-oscillations[8,26]. Coupled to both near-fields, however, the electron spectrum develops a strong asymmetry (lower two panels in Fig. 1**c**) towards energy gain or loss, controlled by the relative phase of both fields (cf. Fig. 1**d**).

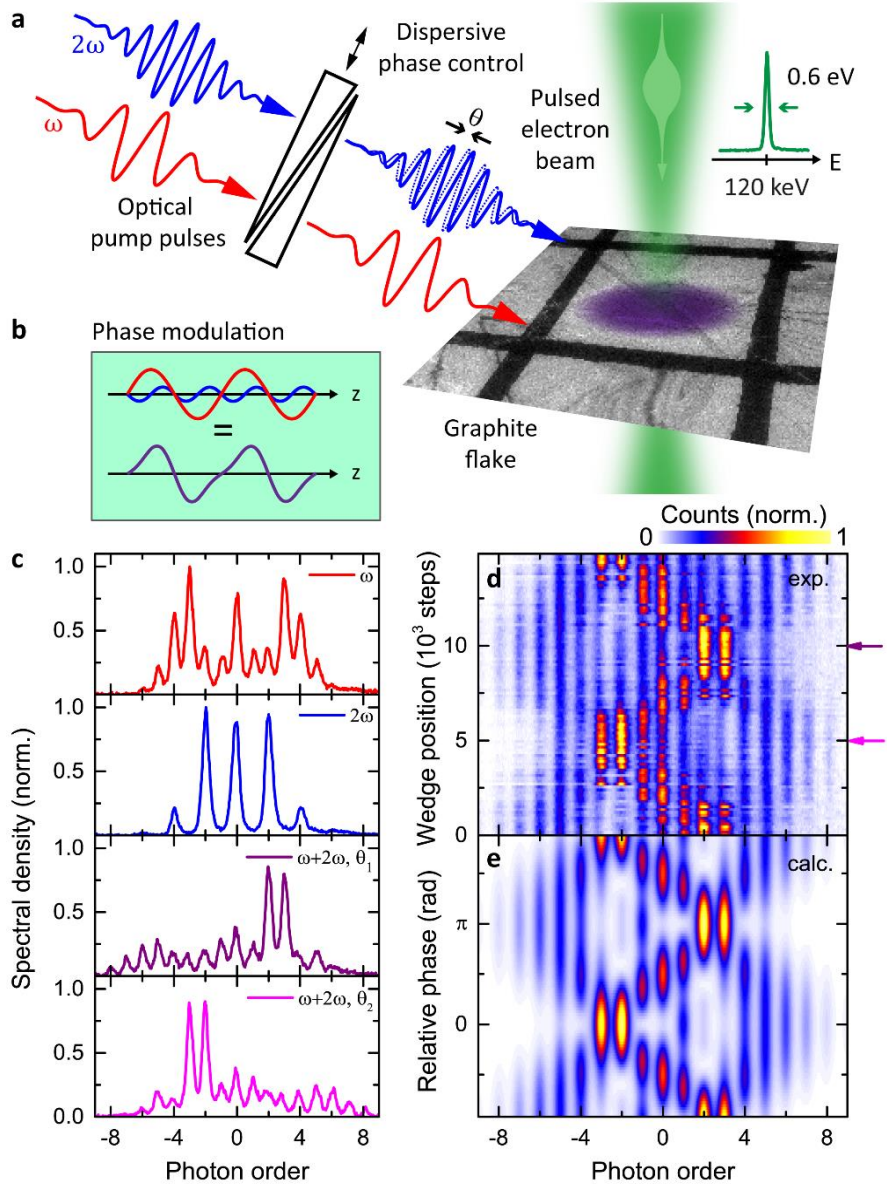

Figure 1: *Experimental scheme.* a) Optical pump pulses at frequencies ω (λ = 800 nm) and 2ω (λ = 400 nm) are spatially and temporally overlapped with a pulsed electron beam on a single-crystalline graphite flake. Fused silica wedges are used to control the relative phase between the laser pulses. An electron-energy-loss spectrometer (EELS) records the electron energy spectrum, which initially exhibits a narrow peak at a central energy of 120 keV and an energy width of 0.6 eV. b) The electron-light-interaction can be described as a phase modulation of the electron wavefunction. For two-color laser fields, the phase modulation becomes non-sinusoidal (purple curve). c) Experimental electron energy spectra recorded for single-color (red and blue curves) and two-color excitation (purple and magenta curves). In the latter case, the spectra are strongly asymmetric and depend on the relative phase of the two colors ($\theta_1=\pi$, $\theta_2=0$). d) The measured spectral shape oscillates back and forth for varying wedge insertion. The spectra in **c** are taken from the positions marked by the purple and magenta arrows. e) The corresponding calculated spectra (Eq. 2) using coupling constants $g_\omega$ = *2.20* and $g_{2\omega}$ = *0.76* for the fundamental and second harmonic, respectively. Contributions from low-loss plasmon bands were subtracted from all spectra. Note that, throughout the paper, the photon order refers to the fundamental frequency ω.

These observations can be rationalized by adapting the theoretical description of inelastic electron-light scattering[29,30,38,26,8] to the present two-color scenario. For interaction with a single light field at frequency $\omega$, the spatial wavefunction of the free-electron quantum state $|\psi\rangle$ experiences a sinusoidal phase modulation in the beam direction[26,30],

$$\psi(z) = \exp\left(2i|g_\omega|\sin\left(\frac{\omega z}{v} + \arg(g_\omega)\right)\right) \cdot \psi_{\text{in}}(z) =: A(g_\omega, \omega) \cdot \psi_{\text{in}}(z). \tag{1}$$

Here, $\psi_{\text{in}}(z)$ denotes the wavefunction of the unperturbed electron quantum state (leaving out dependencies on transverse coordinates for simplicity), $v$ the electron velocity, $z$ the spatial coordinate along the electron trajectory, and $g_\omega$ is a dimensionless coupling constant as defined in Refs.26,30. Equivalently, the quantum state can be written as a coherent superposition of momentum sidebands[26,29,30]. The action of two fields at frequencies $\omega$ and $2\omega$ is now described in terms of two superimposed phase modulations, which for the typically small total energy changes (relative to the initial electron energy) results in the electron quantum state

$$\psi_{\text{out}}(z) = A(g_\omega, \omega) \cdot A(g_{2\omega}, 2\omega) \cdot \psi_{\text{in}}(z), \tag{2}$$

where $g_\omega$ and $g_{2\omega}$ are the two complex coupling constants. Overall, the dual phase modulation is non-sinusoidal (cf. Fig. 1**b**), resulting in the observed asymmetric electron spectra. The phase-dependent experimental spectrograms (Fig. 1**d**) are reproduced by a cycling of the relative phase $\theta = \arg(g_\omega) - \arg(g_{2\omega})/2$ in Eq. (2). A rich variety of tailored quantum states is accessible by variation of the relative phase and amplitudes of such bichromatic fields, and a further design of such momentum state synthesis may be realized by optical pulse-shaping techniques[39].

Multiple phase-controlled interactions at one or more frequencies not only enable the preparation but also the characterization of free-electron quantum states, as we demonstrate in the following. Slightly shifting our perspective on the experimental scenario, we now regard the interaction of the electron with the $2\omega$-field as the preparation of a specific quantum state, described by a density operator $\rho$ to account for the possibility of mixed states, which is then probed by the $\omega$-field. Based on this interpretation, we introduce a new variant of quantum state tomography[40,41] termed "**S**pectral **QU**antum **I**nterference for the **R**egularized **R**econstruction of free-**EL**ectron **S**tates", abbreviated as "SQUIRRELS". As illustrated in Fig. 2**a** and detailed in the Appendix (section 3), SQUIRRELS reconstructs the free-electron density matrix $\rho$ in the longitudinal momentum basis from experimental spectrograms. Specifically, the action of the ω-field on $\rho$, described by a unitary transformation $U$, results in a final quantum state $\rho_{\text{out}}$ that depends on the relative phase $\theta$,

$$\rho_{\text{out}}(\theta) = U(\theta)\rho U^\dagger(\theta) \quad \text{with} \quad \langle N|U(\theta)|M\rangle = e^{i(N-M)\theta} J_{N-M}(2|g_\omega|). \quad (3)$$

Here, the integers $N$ and $M$ label the electron momentum states of the individual photon sidebands (positive/negative for energy gain/loss), and $J_{N-M}$ denotes the Bessel function of the first kind. Note that Eq. (3) generalizes Eq. (1) to mixed states and treats the ω-field as a type of local oscillator, which in the present context is regarded as an ideal phase modulator. The populations $p_{N,\theta} = \langle N|\rho_{\text{out}}(\theta)|N\rangle$ constitute our observables, namely the phase-dependent sideband intensities in the spectrogram (Fig. **2b**). While the diagonal entries of $\rho$, namely the populations $\langle N|\rho|N\rangle$ of the prepared quantum state, can be readily measured in a single-color experiment, the off-diagonal terms or coherences $\langle N|\rho|M\rangle, N \neq M$ initially remain unknown and must be reconstructed from the two-color data $p_{N,\theta}$. In order to obtain the full density matrix $\rho$, we thus use Eq. (3) to solve a linear system of coupled equations, which in mathematical terms is ill-posed. Stable solutions of the resulting (ill-conditioned) matrix equation are achieved by iterated Tikhonov regularization, as detailed in the Appendix (section 3), employing the positive-semidefiniteness of physical density matrices as a constraint on $\rho$. We note that the present scenario is closely related to established techniques for the retrieval of spectral phases of ultrashort and attosecond optical pulses, such as FROG[42] and RABBITT[43]. In the Appendix (section 5), we also apply RABBITT to reconstruct the free-electron quantum state.

Figure 2 presents an exemplary SQUIRRELS reconstruction, in the form of Wigner functions[44] of the intermediate ($\rho$) and final quantum state ($\rho_{out}$). The Wigner function is a quantum-mechanical quasi-probability distribution in phase space that completely describes the quantum state of the electron ensemble, and whose marginal distributions, i.e. integrals along horizontal and vertical axes, correspond to the density distributions of the longitudinal momentum and position, respectively. Negative values of the Wigner function illustrate the non-classical nature[44,45] of the electron quantum state. Albeit being equivalent to the density matrix, the Wigner function provides a more intuitive representation by revealing the sinusoidal momentum modulation (Fig. **2c**) induced by the interaction (Further reconstructions at growing field amplitudes are shown in Suppl. Fig. 5). This sinusoidal shape is complemented by a mirrored sinusoidal feature composed of alternating positive (red) and negative (blue) stripes, such that electron energies corresponding to non-integer photon numbers destructively interfere. The non-sinusoidal momentum modulation of the corresponding final two-color state $\rho_{out}$ is apparent in Fig. 2d.

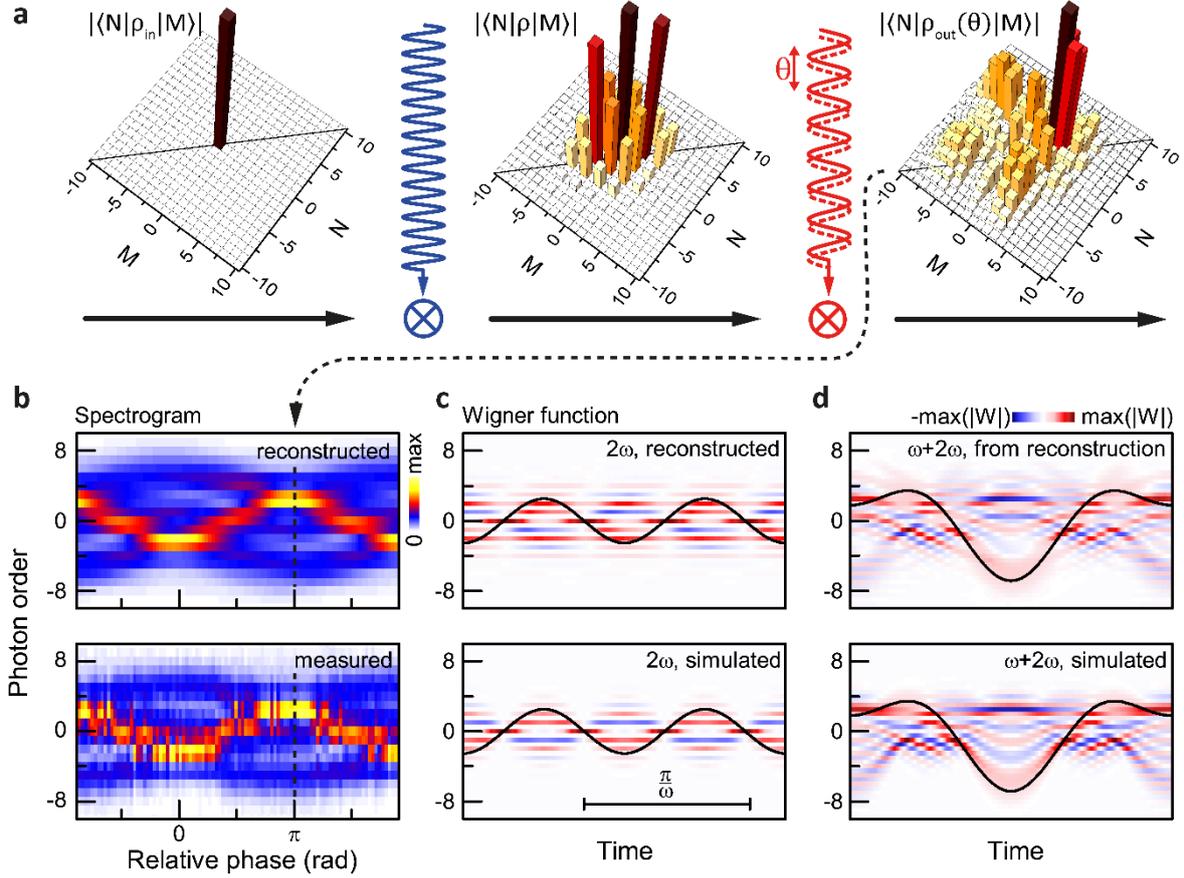

Figure 2: *SQUIRRELS reconstruction of the free-electron quantum state.* a) Reconstructed density matrices and illustration of the underlying tomographic principle: Preparation of the free-electron quantum state with density matrix $\rho$ is obtained by applying a laser pulse at frequency $2\omega$ to the incident quantum state $\rho_{in}$. In a second step, a laser pulse at frequency $\omega$ and relative phase $\theta$ with respect to the first pulse probes the quantum state $\rho$ by transforming it into $\rho_{out}$. Note that only the populations (diagonal elements, marked by the black line) of the density matrices $\rho$ are accessible in the measurement, the coherences (off-diagonal elements) remain unknown. The shown density matrices $\rho$ and $\rho_{out}(\theta=\pi)$ were reconstructed from experimental data. b) Spectrogram containing the phase-dependent populations $\rho_{out}(\theta)$. Upper panel: reconstructed, lower panel: measured. c) The reconstructed and simulated Wigner functions for the single-color quantum state $\rho$ illustrate the sinusoidal phase modulation. d) Corresponding two-color Wigner functions for $\rho_{out}(\theta=\pi)$. The lower panels in c) and d) show model calculations for pure quantum states ($g_\omega$ = 2.16, $g_{2\omega}$ = 0.63). Black solid lines: phase modulation according to Eq. 1 or 2 as guide to the eye.

Instead of employing two-color fields in a single interaction plane, quantum state reconstruction is also possible by sequential actions in separate planes, either by dual or single-color fields. In the following, we implement this concept in two scenarios, namely a µm-sized and a millimeter-sized separation of interaction distances. Figure 3 presents SQUIRRELS applied to a measurement conducted in the geometry introduced in Ref. 8, with a few-micron distance between two phase-locked near-field interactions of the same frequency. Excellent agreement between the reconstructed density matrix (Fig. 3c) and Wigner function (Fig. 3d) with a corresponding simulation (Figs. 3e,f) is found, with only minor loss of phase coherence indicated by damped elements far off the main diagonal.

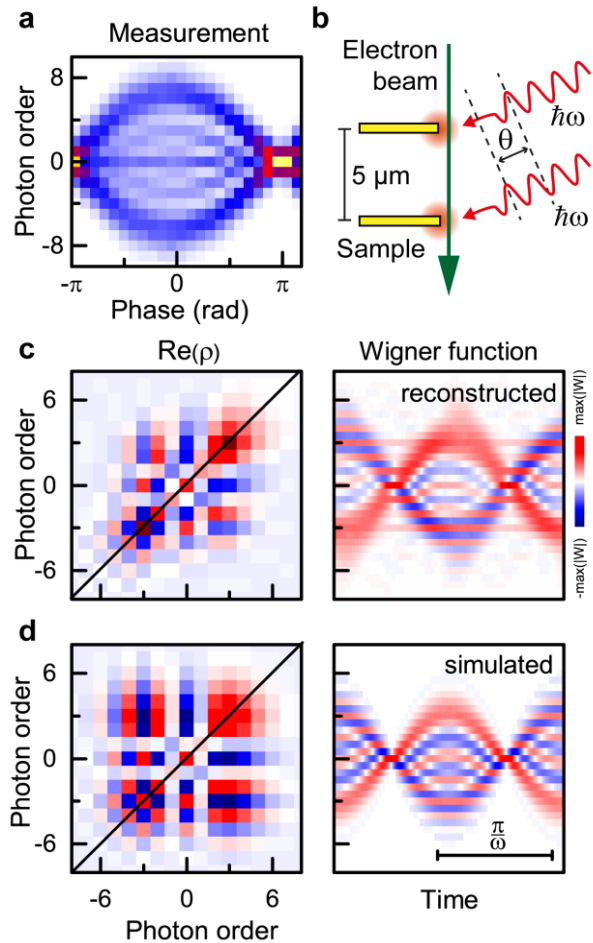

Figure 3: *Application of SQUIRRELS to spatially separated optical near-fields.* a) Experimental spectrogram (data from structure in Ref. 8). b) Sketch of the experimental scenario. c) Reconstructed density matrix (left) and Wigner function (right) of the electron quantum state prepared by the first optical near-field after free-space propagation over a distance of 5 µm. d) Corresponding simulations for a pure state with *g* = 1.97.

We now apply this scheme to experimentally demonstrate the creation of a train of attosecond density spikes, as recently proposed[26]. In the measurements presented in Fig. 4, the distance to the second interaction plane is increased to 1.5 mm. This allows for a dispersive reshaping of the electron density by a shearing of the phase-space distribution, as also utilized in accelerator-based applications of longitudinal beam structuring[46]. In Figure 4**b**, the final spectrum is displayed as a function of relative phase over multiple cycles. Using SQUIRRELS, we retrieve the corresponding sub-cycle electron density structure (Fig. 4**d**), which exhibits a baseline density at 0.27 of the maximum value, and, notably, a train of attosecond peaks of a width of 277 as (root-mean-square or rms; full-width-at-half-maximum: 655 as). Accordingly, the high-quality Wigner function reconstruction (Fig. 4**c**) exhibits a sheared sinusoidal shape, with many fine interference features. From a comparison with model simulations, we estimate that spatial and temporal averaging over different mutual phases in both planes is limited to below 189 mrad (80 as rms, cf. Suppl. Fig. 6). In the present experiments, geometrical constraints limited the dispersive propagation to 1.5 mm, while the shortest attosecond pulses are expected for 2.75 mm propagation for $g_{pump}$ = 3.95. The pronounced attosecond density modulation achieved here enables the nanoscale exploration of optically-driven electronic and valence changes in electron microscopy with sub-cycle, attosecond accuracy. In future experiments, a further reduction in pulse duration to less than 100 as seems feasible, employing optimized propagation distances, field strengths and phase stability. Moreover, also the quantitative reconstruction of isolated attosecond electron pulses will be possible by adapting the approach presented.

In conclusion, we demonstrated the coherent control, quantum state reconstruction and attosecond structuring of free-electron beams. The approach links ultrafast transmission electron microscopy with tools from both attosecond spectroscopy and quantum optics. We envisage the application of this framework in novel quantum measurement schemes in electron microscopy, yielding structural and electronic observables with nanometer spatial and attosecond temporal resolutions, possibly on the level of single quantum systems. Extending the approach to transverse scattering of electrons will establish the programmable, three-dimensional shaping of free-electron wave packets as a basic element of free-electron quantum optics technology.

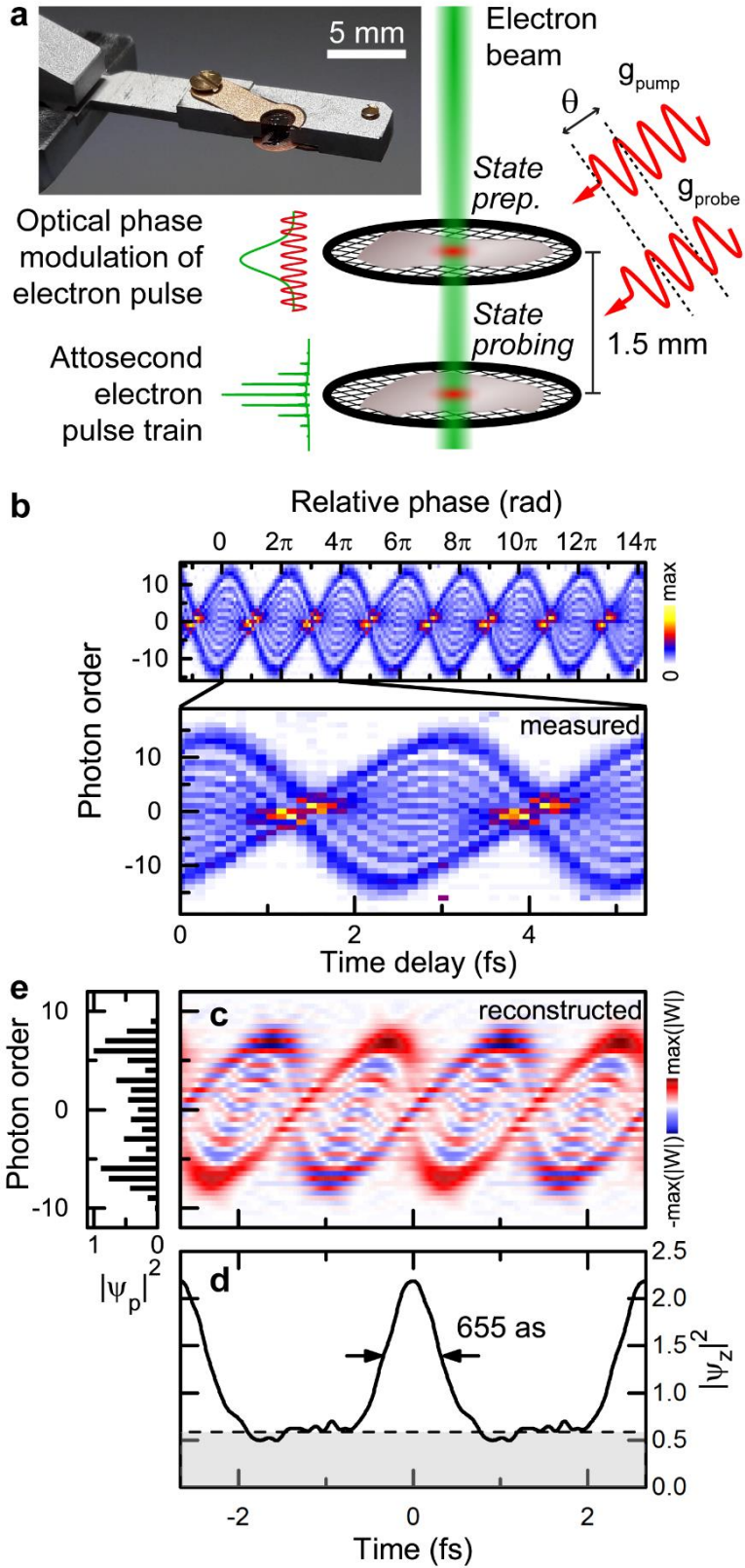

Figure 4: *Experimental demonstration of attosecond electron pulse trains.* a) Sketch of the experimental setup employing two graphite flakes for the preparation (upper plane) and characterization (lower plane) of attosecond electron pulse trains. Inset: Photograph showing the custom-built TEM sample holder. b) Experimental spectrogram recorded over multiple optical cycles and close-up of two cycles. c) The reconstructed Wigner function (using $g_{probe}$ = 3.52) reveals a pronounced shearing due to free-space propagation. d) The temporal projection of the Wigner function exhibits density modulations with a rms pulse duration of 277 as (after baseline subtraction, full-width-at-half-maximum: 655 as). e) Corresponding electron energy spectrum (momentum projection). The results are in excellent agreement with calculations employing pure states (cf. Suppl. Fig. 6).

**Acknowledgements**

We gratefully acknowledge funding by the Deutsche Forschungsgemeinschaft (DFG-SPP-1840 'Quantum Dynamics in Tailored Intense Fields', and DFG-SFB-1073 'Atomic Scale Control of Energy Conversion', project A05), support by the Lower Saxony Ministry of Science and Culture and funding of the instrumentation by the DFG and VolkswagenStiftung. We thank O. Kfir for useful discussions. T.H. would like to thank Anja Fischer (Göttingen) and Kim-Chuan Toh (Singapore) for helpful discussions on semidefinite programming.

# Appendix

## 1. Experimental details

The experiments were performed in an ultrafast transmission electron microscope equipped with a nanoscopic tip emitter, as described in detail in Refs. 8,26,34. Supplementary Figure 1 depicts the optical beam path and the interferometer designs used for the two different sets of measurements. A pulsed laser beam from an amplified fs-laser system (250 kHz repetition rate, 800 nm central wavelength, 50 fs pulse duration) is split in two parts, one of which is frequency-doubled in a β-barium borate (BBO) crystal and focused onto a zirconium-oxide covered tungsten tip to generate a pulsed photoelectron beam (probe beam). For the two-color experiments, an interferometer labelled '**A**' in Suppl. Fig. 1 was set up: The second part of the laser beam (pump beam) is frequency-doubled in another BBO crystal and separated into two beam paths at 800 nm and 400 nm wavelength. The 800-nm and 400-nm pump pulses are stretched to a duration of 2.7 ps and 1.3 ps (cf. Suppl. Fig. 2), by propagation through a 19-cm SF6 and a 10-cm BK7 glass slab, respectively. This ensures laser pulse durations exceeding that of the electron pulse, such that the electrons experience a constant near-field amplitude (see Ref. 26) and the electron-light interaction can be described by a single coupling constant as in Eq. (3), a requirement for the present reconstruction algorithm. The two laser beams at frequencies $\omega$ and $2\omega$ are recombined and focused onto the sample within the TEM chamber (~30 μm spot size) after passing two wedges (fused silica, wedge angle 4°) for precise phase control. The electron beam (~17 nm focus size) and both pump laser beams are spatially and temporally overlapped on a single crystalline graphite flake (about 100 nm thick), obtained by mechanically cleaving a natural graphite single crystal.

To measure the attosecond temporal structuring of the electron density, we implemented a custom TEM holder capable of carrying two TEM grids with single-crystalline graphite flakes, spatially separated by 1 mm. A second interferometer (labelled '**B**' in Suppl. Fig. 1) equipped with a motorized mirror mount in one of the interferometer arms allows for an independent control of the laser focus positions on the top and bottom sample planes. The interferometer is actively stabilized using a 400-nm cw-laser. The electron beam diameter was increased to a ~3 μm focus size to reduce the influence of mutual phase differences between the optical excitation of the top and bottom interaction regions. For both experiments, the resulting electron energy distribution is recorded in an electron spectrometer with 5 s and 40 s integration time, respectively.

## 2. Data analysis

Besides the coherent interaction with the optical near-field, the electron may also interact with the sample itself, e.g. by plasmon excitation, giving rise to a weak, spectrally broad energy-loss feature in the recorded spectra, which was removed from the data. While the energy spectra are recorded with an energy resolution better than the photon energy, we reduce the experimental data to the photon sideband populations for further analysis. To this end, we employ a global fit function consisting of Pseudo-Voigt profiles separated by the photon energy, which are offset by an asymmetric Gaussian describing the plasmon contribution. The obtained sideband amplitudes constitute a reduced form of the spectrograms, which serve as the input to the reconstruction algorithm.

A reliable reconstruction result requires knowledge of the probe pulse coupling constant $g_\omega$, since it is a parameter entering the unitary operator $U$ in the reconstruction algorithm. The value of $g_\omega$ can be obtained in multiple ways: For instance, if an experimental single-color spectrum has been recorded for the same excitation conditions as in the two-color spectrogram, fitting Bessel amplitudes to this single-color spectrum yields $g_\omega$ (see also Appendix section of Ref. 8). Alternatively, the two-color spectrogram can be fitted by Eq. (2), yielding values for both $g_\omega$ and $g_{2\omega}$ corresponding to the pure states which are closest to the experimental conditions. Finally, $g_\omega$ can be obtained with an optimization routine on the SQUIRRELS algorithm which minimizes the discrepancy between the experimental and reconstructed spectrogram under variation of $g_\omega$. All approaches have resulted in very similar values for the coupling constants.

## 3. Description of SQUIRRELS algorithm

Let us consider the electron density matrix reconstruction within the framework of closed quantum systems. In this case, the density operator evolves according to the time-dependent Liouville-von Neumann equation

$$\frac{d\rho}{dt} = -\frac{i}{\hbar}[H + H_{2\omega}(t) + H_\omega(t), \rho(t)], \tag{E1}$$

where H is the Hamiltonian of the electron in the absence of any laser field, and $H_{2\omega}(t)$ and $H_\omega(t)$ describe its interaction with two overlapping quasi-monochromatic laser pulses, $A_{2\omega}(t)\cos(2\omega t)$ and $A_\omega(t)\cos(\omega t+\Theta)$, respectively. As was shown in Refs. 8,26, if the energy transfer during the interaction is small compared with the initial energy of the electron, then $H_{2\omega}(t)$ and $H_\omega(t)$ can be regarded as commuting operators. Consequently, the unitary transformation in the interaction picture can be split into a product of two commuting unitary operators, $U_{2\omega}$ and $U_\omega$, associated with each laser pulse. As a result, the quantum evolution from an initial state $\rho_{in}$ at $t = -\infty$ to a final state $\rho_{out}$ at $t = +\infty$ can be seen as a two-step process passing through an intermediate state. This situation may be illustrated by the diagram

$$\rho_{in} \xrightarrow{U_{2\omega}} \rho \xrightarrow{U_\omega(\theta)} \rho_{out}(\theta).$$

In this diagram, the first action serves as the preparation of a quantum state, with which a second, phase-controlled field interacts. The main difficulty in the determination of a quantum state stems from the lack of knowledge about the coherent (off-diagonal) part of the density matrix in quantum measurements. Here, we show how this information can be retrieved in a series of von Neumann's selective projective measurements[47], where the diagonal elements of $\rho_{out}(\Theta)$ are measured at different phase delays $\Theta$ between the two fields. This provides statistical information necessary for a reconstruction of the unknown off-diagonal elements of the intermediate-state's density matrix ρ.

The second step in the diagram is described, in the interaction picture, by the unitary transformation

$$\rho_{\text{out}}(\theta) = U_\omega(\theta)\rho U_\omega^\dagger(\theta), \qquad U_\omega(\theta) = \mathcal{T}\exp\left(-\frac{i}{\hbar}\int_{-\infty}^{\infty} H_{\omega,int}(t)dt\right), \qquad \text{(E2)}$$

where $\mathcal{T}$ is the time-ordering operator. We use the dagger notation (†) to denote the Hermitian conjugation. In the basis of eigenstates of H, $H|l\rangle = (E_0 + l\hbar\omega)|l\rangle$, $U_\omega(\Theta)$ is given by[8,26,30]

$$\langle k|U_\omega(\theta)|l\rangle = \exp(i(k-l)\theta)J_{k-l}(2|g|), \qquad \text{(E3)}$$

where $J_{k-l}(2|g|)$ is the Bessel function of the first kind, and $g$ is the coupling constant associated with the second laser pulse. The measurement is described by a positive operator-valued measure (POVM) with operators $\Pi_l = |l\rangle\langle l|$ such that the probability for the outcome $l$ to occur in the experiment with a given set of phase delays $\theta \in (0, \pi)$ is given by $p_{l,\theta} = tr[\Pi_l \rho_{\text{out}}(\theta)] = \langle l|\rho_{\text{out}}(\theta)|l\rangle$. Combining this expression with Eq. (E2), we obtain the mapping of the unknown density matrix $\rho$ to the experimental data

$$[T(\rho)]_{l\theta} = p_{l,\theta}, \qquad (E4)$$

where $T$ is a linear operator defined by $[T(\rho)]_{l\theta} := \langle l|U_\omega(\theta)\rho U_\omega^\dagger(\theta)|l\rangle$. Although the Hilbert space is infinite-dimensional, in practice essentially only a finite number of states $m = 2l_{max}+1 \approx \Delta E/\hbar\omega$ is occupied, corresponding to the expected energy width $\Delta E$ of the quantum state $\rho$. Therefore, $T$ is very well approximated by an operator on the finite-dimensional space $\mathcal{X}$ of Hermitian complex matrices $\rho$ with $\rho_{kl} = 0$ if $k$ or $l$ are odd. The latter follows from the fact that only states $|l\rangle$ with even $l$ can couple to $|0\rangle$ due to the second harmonic interaction. $\mathcal{X}$ is naturally equipped with the Hilbert-Schmidt inner product $\langle \rho, \tilde{\rho} \rangle = tr(\rho^\dagger \tilde{\rho})$ and the corresponding norm $\|\rho\|^2 = \langle \rho, \rho\rangle = \sum_{k,l}|\rho_{kl}|^2$.

It turns out that the inverse problem (E4) is ill-posed in the sense that $T$ does not have a bounded inverse with respect to any natural norm, which leads to ill-conditioned finite matrices and implies that noise in the experimental data is strongly amplified by "naïve" matrix inversions. A remedy against ill-posedness is regularization. We use variational or Tikhonov regularization as one of the most well-known and commonly used regularization methods (see, e.g., Ref. 48), since it is very flexible and in particular allows us to incorporate the *a-priori* knowledge that $\rho$ is positive semidefinite as a constraint into the inversion process:

$$\rho_\alpha = \underset{\rho \in \mathcal{X}}{\text{argmin}} \left[ \|T(\rho) - p\|^2 + \alpha \|\rho - \rho^{(0)}\|^2 \right] \text{ subject to } tr(\rho) = 1, \rho \geq 0. \qquad (E5)$$

The penalty term $\alpha\|\rho - \rho^{(0)}\|^2$ with a regularization parameter $\alpha > 0$ and some initial guess $\rho^{(0)}$ (in our case, $\rho^{(0)} = 0$) already restores stability, but the constraint $\rho \geq 0$ has an additional strongly stabilizing effect. Equation (E5) can also be interpreted as a maximum posterior estimator from a Bayesian point of view where the term $\alpha\|\rho - \rho^{(0)}\|^2$ corresponds to the prior[49]. Equation (E5) has the form of a quadratic semidefinite program (SDP)[50], the numerical solution of which will be discussed later.

Often, the approximation error can be reduced by iterating Eq. (E5) in the form

$$\rho_\alpha^{(j+1)} = \underset{\rho \in \mathcal{X}}{\operatorname{argmin}} \left[ \|T(\rho) - p\|^2 + \alpha \|\rho - \rho^{(j)}\|^2 \right] \text{ subject to } tr(\rho) = 1, \rho \geq 0. \tag{E6}$$

This is known as iterated Tikhonov regularization[48] and can also be interpreted as an instance of the proximal point algorithm[51] for minimizing $\|T(\rho) - p_{out}\|^2$ under the constraints $tr(\rho) = 1$ and $\rho \geq 0$. We always performed three iterations of Eq. (E6) since on simulated data we only obtained significant improvements in the first three iterations.

To choose the regularization parameter $\alpha$ in Eq. (6), we use the discrepancy principle[48]

$$\alpha = \sup_{\beta \in \mathcal{A}} \beta, \quad \mathcal{A} = \left\{ \beta : \beta > 0, \left\| T\left(\rho_\beta^{(3)}\right) - p \right\| \leq \tau \delta \right\}. \tag{E7}$$

Classically, $\delta$ denotes a bound on the noise level, i.e. $\|T(\hat{\rho}) - p_{out}\| \leq \delta$ where $\hat{\rho}$ is the true (unknown) density matrix. Since such a bound is difficult to obtain in our case, we chose $\delta = \lim_{\alpha \to 0} \left\| T\left(\rho_\alpha^{(3)}\right) - p \right\|$. We point out that, in our case, $\left\| T\left(\rho_\alpha^{(3)}\right) - p \right\|$ is a monotonically increasing function of $\alpha$, and thus the limit $\delta$ is always non-negative. With this definition, the signal-to-noise ratio $\|\rho_{out}\|/\delta$ takes values between 3.8 and 6.4 for our experimental data sets. With the parameter *τ = 1.01*, the choice of $\alpha$ according to Eq. (E7) yields good results for simulated data in all our experimental settings and plausible results for our experimental data.

We return to Eq. (E5) and discuss an equivalent transformation of the quadratic SDP (E5) into a linear SDP with a quadratic cone constraint[52], which we solve with the help of the open source optimization software SDPT3-4.0[53]. Let ***T*** be a matrix representation of the linear operator *T*, and ***R†R=T†T+αI*** be the Cholesky decomposition with $\mathbf{R} \in \mathbb{C}^{m^2 \times m^2}$. Then $\|T(\rho) - p\|^2 + \alpha \|\rho - \rho^{(0)}\|^2 = \|R(\rho)\|^2 - 2\langle \rho, T^\dagger(p) + \alpha \rho^{(0)} \rangle + C$, where *R* is the operator associated with the matrix ***R***, and *C* is a constant independent of $\rho$. Therefore, the problem (E5) is equivalent to

$$\rho_\alpha = \underset{\rho, t, s}{\operatorname{argmin}} \left[ \frac{t}{2} - \langle \rho, T^\dagger(p) + \alpha \rho^{(0)} \rangle \right] \tag{E8}$$

subject to $t \geq \|s\|^2, s = R(\rho), \quad tr(\rho) = 1, \rho \geq 0.$

The paraboloid $\{(t,s) \in \mathbb{R} \times \mathbb{C}^{m \times m} : t \geq \|s\|^2\}$ can be described as a section of the quadratic cone $\mathcal{K} := \{(u,v,s) \in \mathbb{R}^2 \times \mathbb{C}^{m \times m} : u^2 \geq \|s\|^2 + v^2\}$ by a change of variables $t = 2v - 1 = 2u + 1$. This leads to the equivalent linear semidefinite program (SDP)

$$\rho_\alpha = \underset{\rho \in \mathcal{X}, (u,v,s) \in \mathcal{K}}{\mathrm{argmin}} \left[ v - \langle \rho, T^\dagger(p) + \alpha \rho^{(0)} \rangle \right] \tag{E9}$$

subject to $s = R(\rho), tr(\rho) = 1, \rho \geq 0,$

which was solved by SDPT3-4.0 using an infeasible primal-dual interior point method. Actually, this software cannot treat complex SDPs directly, but supports the conversion of complex SDPs into equivalent real SDPs with matrices of double size.

## 4. Performance of reconstruction

We would like to comment on how to choose the probe strength $g_\omega$ for optimal reconstruction results. While our reconstruction method could in principle be applied for arbitrarily small probe strengths, it is advised to employ values $g_\omega \cong 2g_{2\omega}$, as we will discuss in the following. To test the algorithm performance, we conducted numerical experiments in which we added Poisson noise to synthetic spectrograms calculated from pure-state density matrices. The numerical experiments were repeated for six different values of $g_{2\omega}$ to exclude a dependence on the absolute pump strength. Suppl. Fig. 3 illustrates the main findings: The reconstruction error decreases exponentially with the probe-pump ratio, until a noise level dependent minimal value is reached around $g_\omega/g_{2\omega} \approx 3.5$. This illustrates severe ill-posedness of the inverse problem (E4) for small values of $g_\omega/g_{2\omega}$ corresponding to an exponential decay of the singular values of $T$. We observed numerically that the condition number of discrete representations of $T$ increases exponentially as $g_\omega/g_{2\omega} \to 0$. If $g_\omega/g_{2\omega}$ is increased beyond 3.5, the reconstruction results slowly deteriorate. For $g_\omega/g_{2\omega} = 2$, the respective single-color electron energy spectra have the same absolute energy width (the factor of two results from the ratio of the probe and pump photon energies). Consequently, all sidebands are being interfered with each other, and information about the corresponding coherences is directly encoded in the spectrogram. If $g_\omega$ is small, however, higher-order off-diagonals in the reconstructed density matrix are significantly underestimated, especially for highly noisy data (cf. Suppl. Fig. 3**b**). Mixed states arising from an incoherent average of relative phases between pump and probe pulse would be described by a similar density matrix, such that pure states with noisy spectrograms and true mixed states are indistinguishable for small $g_\omega$. Hence, the probe coupling strength $g_\omega$ should preferably be chosen about two to four times the pump coupling strength $g_{2\omega}$, and noise contributions must be kept below a tolerable level.

## 5. Application of RABBITT

In this Section, we show that a technique known as RABBITT, which stands for "reconstruction of attosecond beating by interference of two-photon transitions" and was invented to measure the relative phases of two neighboring sidebands in high harmonic generation[43], can be adapted to our experimental scenario. To this end, we consider the case where the coupling to the $\omega$-field is small enough to only populate the first-order sidebands $N = \pm 1$, i.e. $|g_\omega| < 0.5$. A pure quantum state prepared by the $2\omega$-field can be written as

$$|\psi\rangle = \sum_{N \text{ even}} c_N |N\rangle \text{ with } c_N = e^{i\frac{N}{2}\arg(g)} J_{\frac{N}{2}}(2|g|) = |c_N| e^{i\varphi_N}, \quad (E10)$$

where $\langle z|N\rangle = e^{ik_N z} = e^{i(k_0 + \Delta k)z}$ is a plane wave with an electron momentum shifted from its initial value $\hbar k_0$ by $\Delta k = N\hbar\omega/v$. The magnitude $|c_N|$ of the sideband amplitudes is readily calculated from the measured spectrogram, while the sideband phases $\varphi_N$ are not directly accessible. In the presence of the weak $\omega$-field, the energy spectrum of the quantum state is only slightly perturbed, but odd-order sidebands are occupied (Suppl. Fig. 4**a**). The population of these intermediate energy levels is governed by interference between the two adjacent sidebands, and is explicitly given by

$$|a_N(\theta)|^2 = J_1(2|g|)^2 \left[ |c_{N-1}|^2 + |c_{N+1}|^2 \right. \quad (E11)$$
$$\left. + 2|c_{N-1}||c_{N+1}| \cos(2\theta + \pi + \varphi_{N+1} - \varphi_{N-1}) \right], N \text{ odd},$$

where $\theta$ is the relative phase between the two laser fields. According to Eq. (E11), the populations of the odd-order sidebands oscillate in a cosine-fashion upon variation of $\theta$, which is clearly visible in the experimental spectrogram (Suppl. Fig. 4**b**). The phase offset in oscillations from different orders encodes the phase difference $\varphi_{N+1} - \varphi_{N-1}$ between two neighboring energy levels, which can thus be obtained from a fit of cosine functions to the experimental sideband intensities. Note that in contrast to the common RABBITT scheme, here, the electrons undergo free-free instead of bound-free transitions, so that atomic phases naturally do not occur and do not have to be accounted for.

The retrieved sideband phases (red squares, Suppl. Fig. 4**c**) are in good agreement with the values expected from Eq. (E10) (black circles). There are, however, two drawbacks in the RABBITT-approach. The first issue concerns experimental uncertainties: The sideband phases are retrieved by adding up phase differences, such that experimental errors cumulate in the higher orders. To overcome this issue, in SQUIRRELS, we employ stronger probe pulses that couple several (ideally all) sidebands to each other. Consequentially, Eq. (E11) is no longer valid, and new algorithms such as SQUIRRELS are required to recover the sideband phases from spectrograms. The second issue concerns the scope of the RABBITT method: Equation (E10) implies a pure quantum state, which generally may not be the case. Pure state (i.e., fully coherent) descriptions may for instance severely underestimate the retrieved pulse durations in ultrashort-pulse characterization methods using partially coherent beams, as discussed in Ref. 54. Our SQUIRRELS method includes the possibility of mixed states, which are generally closer to experimental scenarios, and is thus more widely applicable.

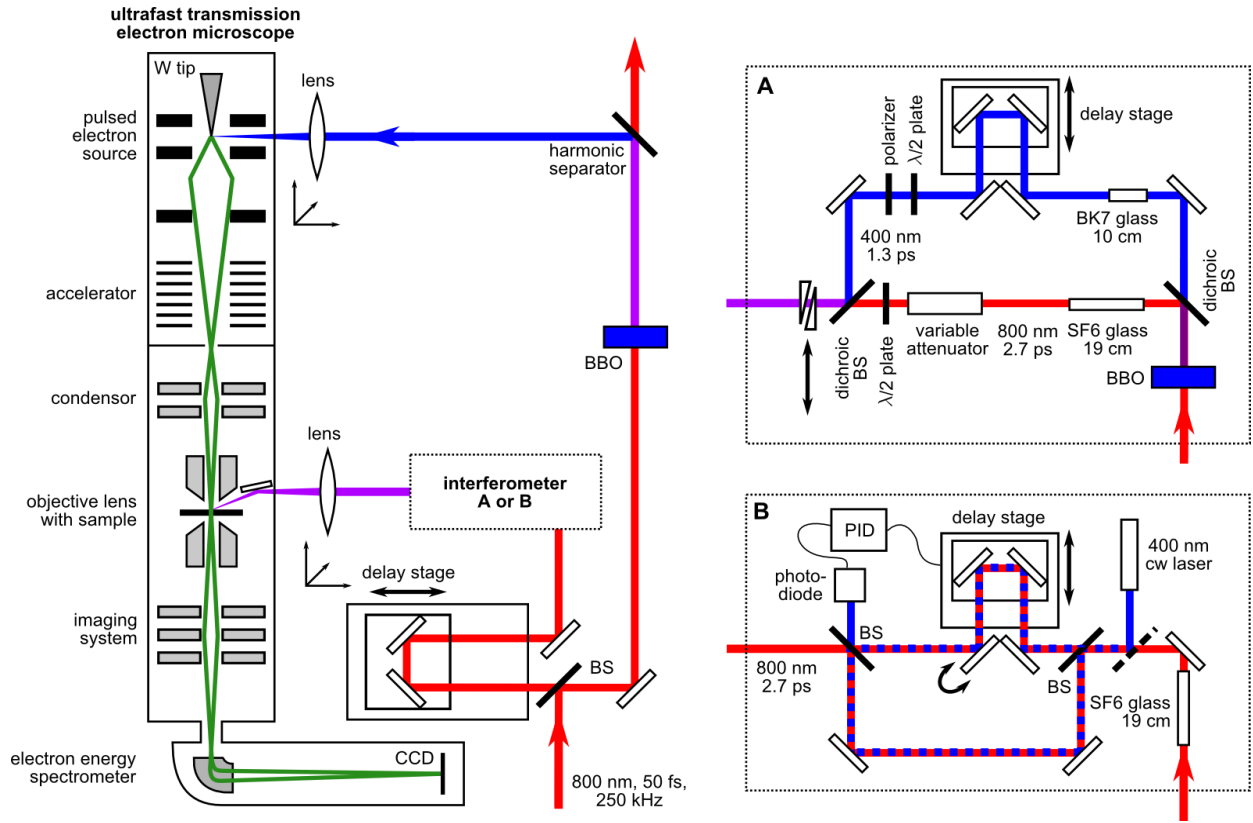

Supplementary Figure 1: Experimental setup. The electron pulses are generated by single-photon photoemission from a heated ZrO/W Schottky-field-emitter using laser pulses at 50-fs pulse duration, frequency doubled to 400-nm wavelength in a BBO crystal. Part of the same laser beam is used for sample excitation to ensure synchronization between the laser-pump and electron-probe pulses. For two-color excitation (interferometer **A**), this part of the beam is further split into two parts, one of which is also frequency-doubled. The linear polarization state as well as the laser intensity can be individually adjusted for both colors. After beam recombination, the relative phase between the two pulses is controlled with fused-silica wedges. For the spatially-separated structure, interferometer **B** is used. A motorized mirror mount in one of the two beam paths allows to create two spatially-separated laser foci within the UTEM. The interferometer is stabilized by a feedback loop (PID control).

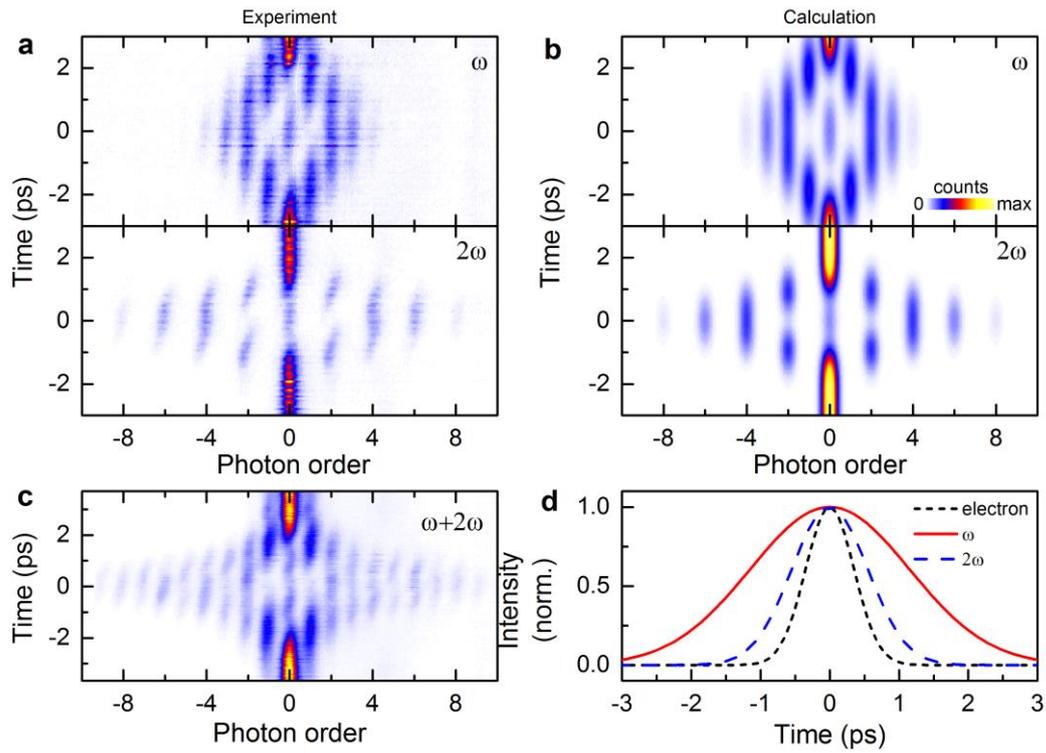

Supplementary Figure 2: *Electron-photon cross-correlation*. a) Measured electron energy spectra as a function of the time-delay between the electron and laser pulses at frequency ω (upper panel) and 2ω (lower panel). b) Corresponding calculations employing Eq. (21) from Ref. 30. c) Electron-photon cross-correlation for two-color excitation to confirm optimized temporal overlap between both laser pulses. d) Intensity envelopes of the three pulses involved, used for the calculation shown in (b). Retrieved pulse durations (FWHM of intensity): 820 fs (electron pulse), 2.7 ps (ω pulse), 1.3 ps (2ω pulse). Electron pulse chirp is not included in the calculation, so that the experimentally observed tilt of the photon sidebands[55] is not reproduced.

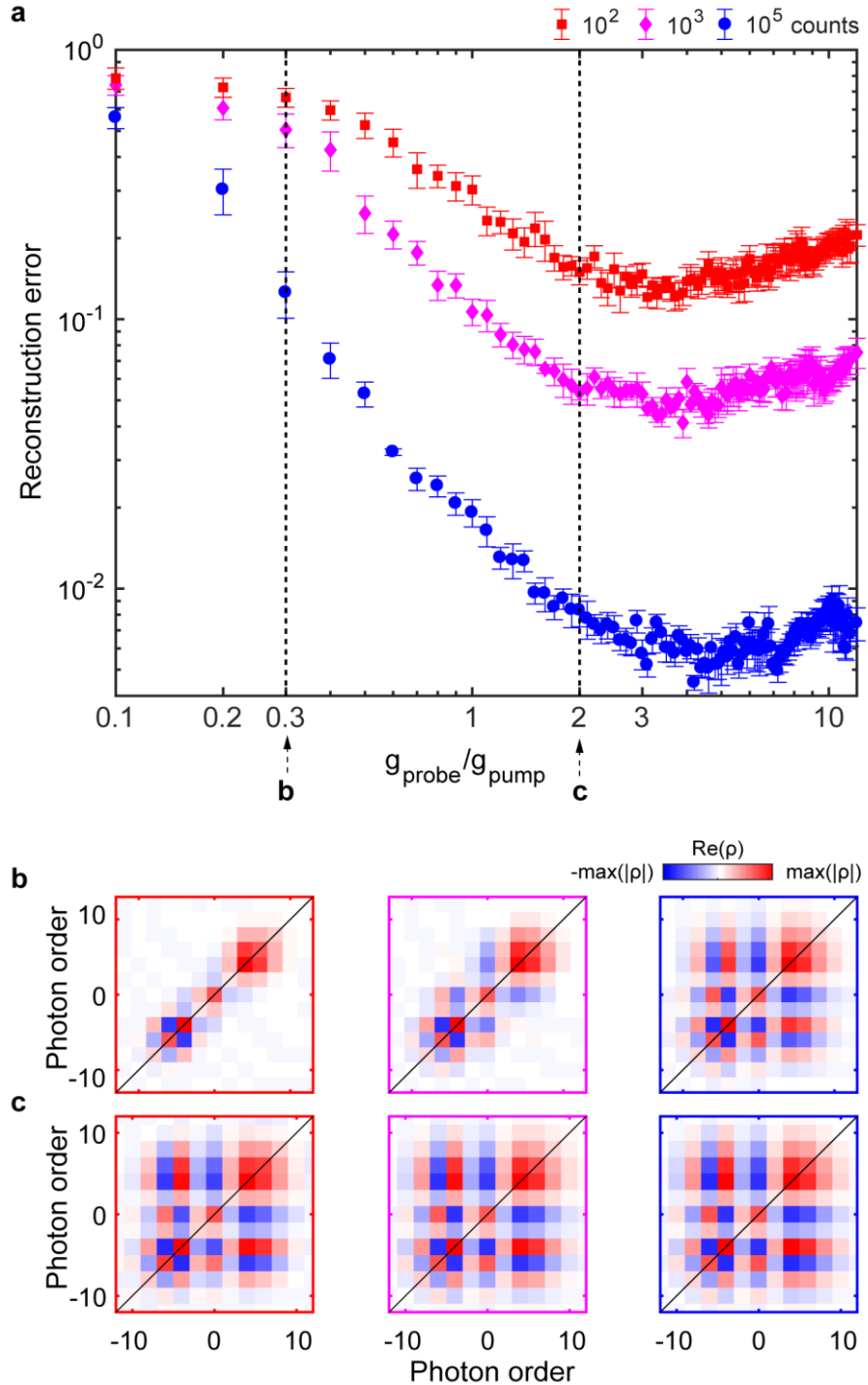

Supplementary Figure 3: *Algorithm performance for noisy synthetic data*. a) We applied the reconstruction algorithm to synthetic spectrograms with different degrees of Poisson noise, i.e., spectra for different numbers of counts per spectrum. The reconstruction error $\|\rho - \hat{\rho}\|_{\mathrm{Fro}}$ decreases with increasing ratio of the probe and pump coupling constant, until it reaches a noise-dependent minimum, followed by a slow increase of the error for even larger ratios. Best reconstruction results are obtained for probe-pump-ratios around three to four. Error bars correspond to the standard deviation of an average over six values for $g_{2\omega}$. b),c) Reconstructed density matrices for decreasing Poisson noise (from left to right) with ratio $g_\omega/g_{2\omega}$ = 0.3 (b) and 2 (c). The reconstruction significantly improves with smaller noise levels and larger ratios $g_\omega/g_{2\omega}$. Pump coupling strength $g_{2\omega}$ = 1.73.

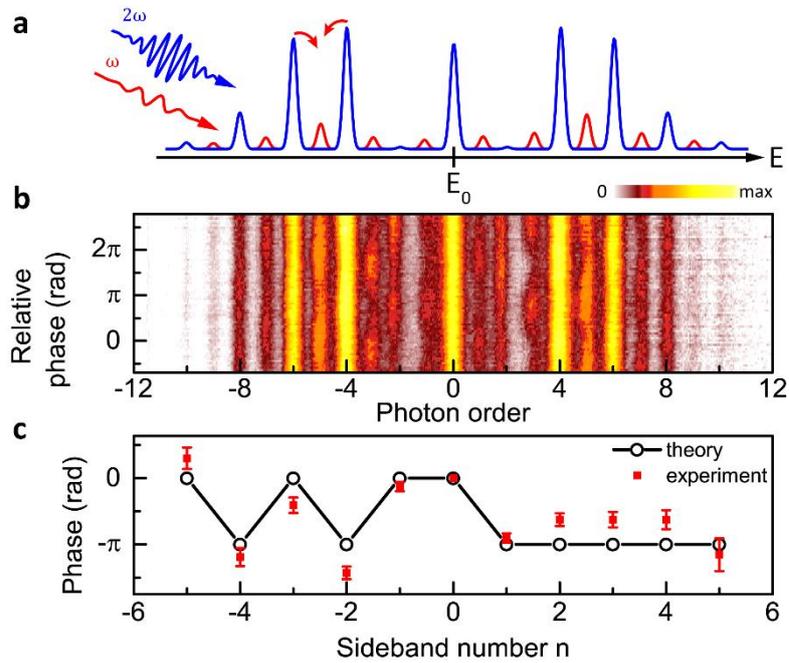

Supplementary Figure 4: *Application of RABBITT to obtain the electron quantum state.* a) Illustration of the underlying principle: A weak probe pulse ($g_\omega$ = 0.13) populates intermediate energy levels (red) in the electron energy spectrum (blue) of the free-electron quantum state as prepared by coherent interaction with the 2ω pulse ($g_{2\omega}$ = 1.85). b) Experimental spectrogram obtained by varying the relative phase of the two-color excitation. The phase-dependent populations of the odd order sidebands exhibit a cosine modulation, whose phase offset encodes the phase difference between two adjacent sidebands. c) The phases of the sideband amplitudes (solid red squares) retrieved from the experimental spectrogram are in good agreement with the values expected from theory (open black circles).

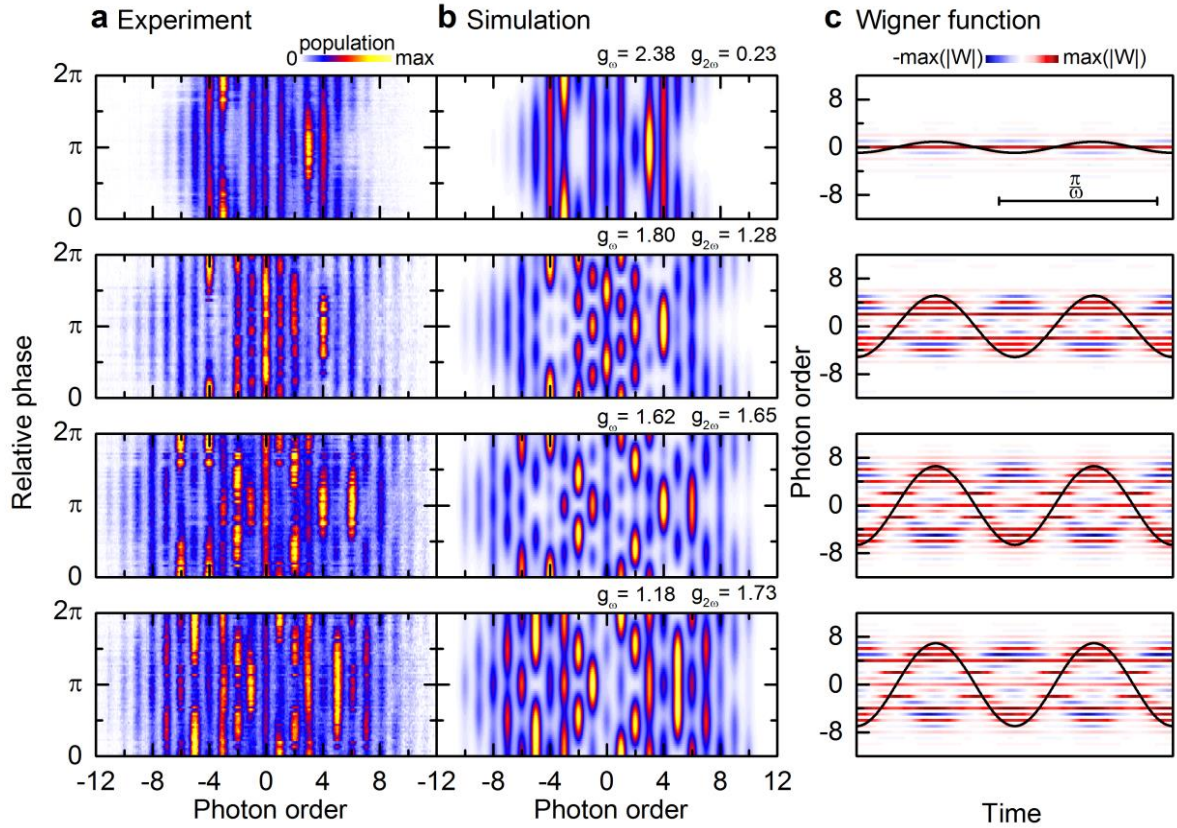

Supplementary Figure 5: *Experimental and calculated spectrograms and corresponding reconstructed Wigner functions.* a) Measured spectrograms after subtraction of the low-loss plasmon band with the full spectral resolution provided by the spectrometer. b) Calculations employing coupling constants as given in the figure reproduce well the prominent phase-dependent spectral features, while minor differences are attributed to phase averaging effects not accounted for in Eq. (2). c) Wigner function reconstructed from experimental spectrograms. The increase of the coupling constant $g_{2\omega}$ from top to bottom is reflected in a growing amplitude of the sinusoidal phase modulation. Black solid lines according to Eq. (1) serve as guide to the eye.

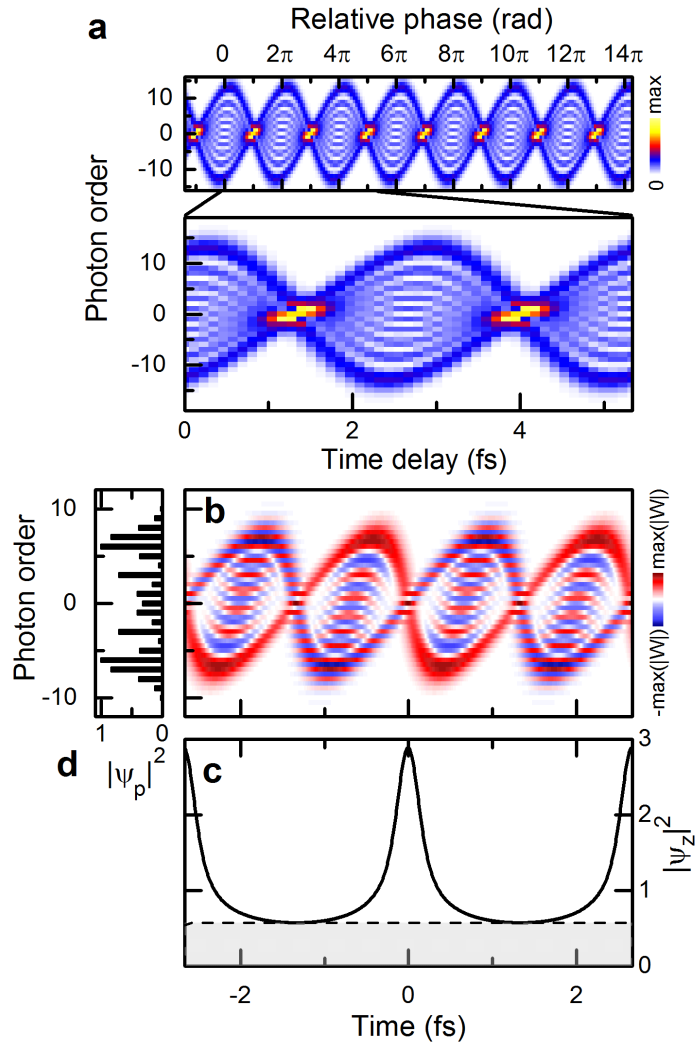

Supplementary Figure 6: *Simulation of attosecond temporal reshaping* a) Simulated spectrogram assuming a pure state with $g_{pump}$ = 3.95 and $g_{probe}$ = 3.52, including a small timing jitter of 80 as (3% of the optical period). These parameters correspond to the experimental values in Fig. 4. b) Corresponding Wigner function. c) The temporal projection of the Wigner function exhibits density modulations with a r.m.s. pulse duration of 296 as (after baseline subtraction, FWHM = 531 as). d) Corresponding electron energy spectrum (momentum projection).